\documentclass[a4paper,reqno,12pt,draft]{article}
\usepackage{amssymb,euscript,bbold}
\newcommand{\be}{\begin{equation}}
\newcommand{\ee}{\end{equation}}
\newcommand{\ba}{\begin{eqnarray}}
\newcommand{\ea}{\end{eqnarray}}

\newcommand{\ft}{\footnote}

\begin{document}
\input{epsf}

\begin{flushright}
RUNHETC-2002-52
\end{flushright}
\begin{flushright}
\end{flushright}
\begin{center}
\Large{\sc A Moduli Fixing Mechanism in $M$ theory.}\\
\bigskip
{\sc B.S. Acharya}\ft{bacharya@physics.rutgers.edu}\\
\smallskip\large
{\sf Department of Physics,\\
Rutgers University,\\ 126 Freylinghuysen Road,\\ NJ 08854-0849.}

\end{center}
\bigskip
\begin{center}
{\bf {\sc Abstract}}
\end{center}
We study $M$ theory compactifications on manifolds of $G_2$-holonomy
with gauge and matter fields supported at singularities. We show that,
under certain topological conditions, the combination of background
$G$-flux and background fields at the singularities induces a potential for
the moduli with an isolated minimum. The theory in the minimum is
supersymmetric and has a negative cosmological constant in the simplest case.
In a more realistic scenario, we find that the fundamental scale is around 10 Tev and the
heirarchy between the four dimensional Planck and electroweak scales may be explained
by the value of a topological invariant. Hyperbolic three-manifolds enter the discussion in an interesting way.

\newpage

\Large
\noindent
{\bf {\sf 1. Introduction.}}
\normalsize
\bigskip

One of the traditional problems which has plagued
string theory and subsequently $M$ theory is the lack of predictivity of
the low energy effective action.
There are apparently many vacua of the theory. For example, in vacua with
compact extra dimensions,
there are
many ways to choose the compactification
data. Each such choice describes different physics and most of these
choices do not seem to describe anything resembling physics as we know it.
At present we know of no compelling
reason why the theory might favour universes with three spatial dimensions
let alone a local symmetry group $SU(3) \times SU(2) \times U(1)$ with
the elementary particles living in three complex representations. On the
other hand, we presently dont know of any other non-anthropic explanations
of these facts either, so string theory is no better than any other
theory in this regard.

A simple manifestation of the vacuum degeneracy/predictivity problem is that
string and $M$ theory compactifications have moduli.
At the classical level this is due to
the fact that the low energy effective potential has many flat directions
and the values of scalar fields which determine the size and shape of the
extra dimensions remain undetermined by the theory.
At the quantum level
there is a similar sort of problem in that non-perturbative effects such as instantons
appear to generate a potential which pushes the fields off to extreme weak
coupling and large volume \cite{ds}. In this regime, the potential is
approximately flat again.

Naively one might think that
the runaway potential might be useful for inflation or quintessence
but the non-perturbative potentials that  arise this way apparently
do not inflate \cite{stein}.

In this paper we will study $M$ theory
compactifications to four dimensions on manifolds of $G_2$-holonomy.
Many such compactifications arise as
a certain limit of Calabi-Yau compactifications of
the heterotic string.
Calabi-Yau compactifications have
traditionally been regarded as the most promising
for obtaining realistic particle physics from string theory.
The limit we will study is not very well
approximated by weakly coupled heterotic strings and is much better described
by $M$ theory on a compact manifold $X$ of $G_2$-holonomy.
Note that $X$
must have certain kinds of singularities in order for
the low energy dynamics to be described
by a non-Abelian gauge theory
coupled to chiral fermions. The singularities leading
to non-Abelian gauge groups were first studied in \cite{bsa1}
and in more detail in \cite{bsa2, amv, aw} whereas singularities
which support chiral fermions have been studied in \cite{aw, ew1,ew2,berg}.
Applications of these studies to more detailed questions of phenomenology
have been made in \cite{ew3,ew4}.

In particular,
in order to obtain non-Abelian local symmetries, the $G_2$-manifold $X$
contains a three manifold $Q$ of orbifold singularities \cite{bsa1}.
Near $Q$, $X$ looks like ${\bf R^4}/{\Gamma}\times Q$.
Hence, $X$ has an orbifold singularity along $Q$.
$\Gamma$ is a finite subgroup of
$SU(2)$ acting as a subgroup of $SO(4)$ on ${\bf R^4}$.
Quarks and leptons
emerge from additional
singularities of $X$ through which $Q$ passes.

Up until relatively recently,
most studies of string compactification focussed on vacua
without fluxes. Typically the metric (as well as non-Abelian gauge fields in the case
of heterotic strings)
is the only non-zero background field and one studies the theory around
a minimum of the classical potential.  Our comments above on flat directions in the
classical potential apply to these cases.
More recently however, attention has been
devoted to the study of compactifications
with flux.
These are compactifications in which the
various $p$-form gauge fields present in the theory have been activated: the
theory is studied in the presence of non-zero electric and/or magnetic fields.
Since the kinetic energy of the $p$-form field depends upon the
spacetime metric, which in turn depends upon the moduli, flux compactifications
lead to a potential for the moduli. Moduli stabilisation due to fluxes has been studied
in a variety of papers $[12]-[30]$.
An open question is can this approach lead to a potential
which fixes all the moduli in a compactification to four dimensions?

In the context of $G_2$-compactifications, the flux induced potential is positive
\cite{beasley} and so
also suffers from a runaway problem. Since most of the interesting physics of $G_2$-compactifications
involves the fields localised at singularities
a simple question that one can ask is
can this runaway be stabilised by turning on the fields supported at singularities of the
$G_2$-manifold? We will study this question in this paper.

We will be interested in the sorts of singular
$G_2$-manifolds that give rise to Yang-Mills theory and chiral fermions and will study the
effect of turning on these fields which are supported at the singularities. We will show that,
as long as a certain topological invariant, $c_2$, is non-zero and suitably large, that
the combination of both fluxes and the fields supported at singularities is enough to fix the
all the moduli of the $G_2$-manifold.

The size of the $G_2$-manifold in the minimum of the potential is large if $c_2$ is large,
so that the approximation to the potential that we use is valid. Thus, if this is indeed the case
then these minima produce four dimensional vacua of $M$ theory with no undetermined parameters.
All masses and couplings are in principle calculable by the theory, although this is difficult
in practice.

In the next section we describe the equations describing the absolute minimum of the supergravity
potential and explain how they can be solved and why the solution is isolated. An important role is
played by a certain complex Chern-Simons invariant $c_1 + i c_2$ which gives an additive contribution to the
superpotential. We show that $c_2$ is non-zero and can be large when $Q$ is a hyperbolic three manifold.

The moduli fixing mechanism presented here involves only the gauge fields along 3-manifolds $Q$ (and their
superpartners) and as such can be applied to {\it non-realistic}
$G_2$-compactifications which do not include quarks and
leptons. In this case, the vacuum we describe is supersymmetric and has a negative cosmological constant
$\Lambda_0$.
In the concluding section, we consider the possibility that
the moduli fixing mechanism is useful in a more realistic
context. For this it is necessary to assume that
the singularities of $X$
support also a supersymmetric standard model {\it and} a supersymmetry breaking
sector. Since by assumption supersymmetry is broken, particle physics interactions will generate a contribution
$\Lambda_1$ to the cosmological constant and the main idea is to balance $\Lambda_0$ against $\Lambda_1$. This is
necessary if one is to obtain (at the very least) a positive cosmological constant.

In this scenario, we find that the fundamental Planck scale is 10 Tev (or smaller) and that the much higher
four dimensional Planck scale emerges because the topological invariant $c_2$ is large. Thus, the heirarchy
of scales that we observe between particle physics and gravity in four dimensions might be explained in these
models by the fact that a topological invariant takes the value that it does.

\bigskip
\Large
\noindent
{\bf {\sf 2. Vacuum Equations for Moduli.}}
\normalsize
\bigskip

$M$ theory on a compact $G_2$-manifold $X$ has a low energy description in terms of a minimally
supersymmetric supergravity theory in four dimensions. This theory is coupled to $b_3 (X)$ moduli
superfields. If $X$ contains a 3-manifold $Q$ along which there is an orbifold singularity then
there will also be non-Abelian gauge fields. Note that there could be several different three manifolds $Q_a$
of orbifold singularities so there can be more than one gauge group factor.
If $Q_a$ passes through certain additional
conical singularities there can be chiral superfields in a complex representation of
the gauge group associated to $Q_a$.

For the purposes of explaining the simple mechanism which fixes the moduli we will only
require that there is a single gauge group and therefore a single $Q$. We wont need to talk about
the matter fields. The classical solution to the low energy equations of motion is a solution of
eleven dimensional supergravity in which the only non-zero field is the 11-metric. This is a product
of the $G_2$-holonomy metric on $X$ and flat four dimensional Minkowski spacetime. Since the
superpotential is exactly zero in this classical low energy limit, the potential for the moduli fields
is also exactly zero.

We will ask the simple question: what happens to the theory if we turn on a background four-form
field $G$ and a background gauge field $A$ along $Q$ ? In the case with $A=0$, this question
has already been studied. The answer is that $G$ induces a superpotential which is linear in the
moduli superfields. In this case, the resulting potential for the moduli is positive and runs off to
zero at infinity \cite{beasley}. This corresponds to the large volume limit.

As we will explain momentarily, turning on the gauge field and its bosonic superpartners leads to a modification
of the superpotential, which in the simplest non-trivial possibility is via the addition of a  constant, $c_1 + ic_2$.
This constant, or rather its imaginary part, has a dramatic effect on the potential: its presence implies
a supersymmetric minimum with negative cosmologial constant. Moreover, all the moduli fields are fixed
in the minimum.
In the remainder of this section we
will explain these results in more detail. In order to do that we need to provide some basic background
material. A more detailed explanation of some of the mathematics may be found in \cite{dominic}.

\bigskip
\large
\noindent
{\bf {\sf 2.1. Kaluza-Klein on $G_2$-manifolds.}}
\normalsize
\bigskip

A manifold of $G_2$-holonomy always comes equipped with a 3-form $\varphi$ which is covariantly
constant (parallel) in the $G_2$-holonomy metric. The existence
of $\varphi$ is equivalent to the existence of a $G_2$-holonomy metric.
Given a fixed $G_2$-holonomy metric on a compact 7-manifold $X$, a new
metric with the same holonomy can be constructed by perturbing $\varphi$ by any small harmonic
three-form. Thus the moduli space of compact $G_2$-manifolds is locally isomorphic to a small neighbourhood
of a point in $H^3 (X, {\bf R})$. There are thus $b_3 (X)$ moduli fields in the four dimensional effective action.
These combine with harmonic fluctuations of the 3-form potential $C$ of low energy $M$ theory to give
$b_3 (X)$ complex scalars which become the lowest components of scalar superfields in four dimensions.

In more detail, let $\phi_j (x)$ $j=1,...b_3 $ be a basis for $H^3 (X, {\bf R})$. Then the Kaluza-Klein harmonic
ansatz is
\be
C + i \varphi = \Sigma_{j} Z_j (y) \phi_j (x)
\ee
The $Z_j(y) $'s are complex scalar fields in four dimensions and $y$ denotes coordinates of four dimensional spacetime.
We define the real and imaginary parts of the complex scalars by
\be
Z_j = {t_j } + i s_j
\ee
The $t_j$'s are axions because of $C$-field gauge transformations.
The $s_j$'s are the moduli fields
whose vacuum expectation values determine the size and shape of the $G_2$-holonomy metric on $X$.
This is because changing the $s_j$'s changes $\varphi$ and therefore also the metric on $X$.

As argued in \cite{gukov,bobill}    and proven in \cite{beasley},
turning on a background field strength $G$ for $C$ induces the following
superpotential into the four dimensional theory
\be
W = {1 \over 8\pi^2} \int_X ({C \over 2} + i\varphi )\wedge G
\ee

Note that as explained in \cite{beasley}the real part
of $W$ is classically an angular variable of period 
${1 \over 2}$.
In the quantum theory, to account for the effects of $G$-flux
through the four-dimensional spacetime, we should accept
all possible values of ${\bf Re}W$ which differ by ${N \over 2}$, for all integers $N$. Although important, the vacuum states distinguished by $N$ will not figure much in what follows.

Since $G$ does not depend upon the metric and is harmonic and quantised, we may expand it in terms
of a basis of harmonic 4-forms $\rho_j$ which are dual to the $\phi_j$. This means that the
superpotential can be written as
\be
W = {1 \over 8\pi^2}\Sigma_k  Z_k G_k
\ee
and is therefore linear in the moduli superfields with coefficients
which are the flux quanta\ft{The quanta $G_k$ are
integer multiples of $2\pi$.}.

Finally, since we will be interested in general supersymmetric vacua,
we will require the Kahler potential and its derivatives. The Kahler
potential depends only on the $s_k$'s and not on the axions. It is given by \cite{jeff,pap,beasley}
\be
K = -3ln(a\int_X \varphi \wedge *\varphi)
\ee
where $a $ is ${1 \over 14\pi^2} $ and $*\varphi$ is the 4-form which is Hodge dual of $\varphi$ in the
metric on $X$. Note that since $*$ depends nonlinearly on the metric, $e^K$ is a nonlinear function of the
moduli. The first derivative of the Kahler potential with respect to the $Z_k$ is \cite{dominic, beasley}
\be
{\partial K \over \partial Z_k} = {7 \over 2}i {\int_X \phi_k \wedge *\varphi \over \int_X \varphi \wedge *\varphi}
\ee

\bigskip
\large
\noindent
{\bf {\sf 2.2. Superpotential for Gauge Fields.}}
\normalsize
\bigskip

In this subsection we will derive the contribution to the superpotential from the gauge fields and
their superpartners which are localised along $Q$ (times four dimensional spacetime $M^{3,1}$).
When the eleven dimensional spacetime has an $A$, $D$ or $E$ orbifold singularity along a 7-manifold
$Y$
such as $Q\times M^{3,1}$, as long as $Y$ is large compared to the Planck length, there
is a description of the $M$ theory physics near $Y$ in terms of seven dimensional super Yang-Mills
theory on $Y$ with an $A$, $D$, $E$ gauge group $G$ determined by the orbifold singularity \cite{bsa1}.
The bosonic fields of 7d super Yang-Mills in flat space are a gauge field $A$ and three scalars $B$.
In our context, the Yang-Mills
theory is on a curved spacetime, but because
$Q$ is a supersymmetric three cycle in a manifold of $G_2$-holonomy
the classical description of the physics for large $Q$ is still supersymmetric even though the spacetime is
curved\ft{See \cite{bsa1} for a detailed discussion of supersymmetry and a description of the low energy physics
in this context.}.
Of the sixteen supersymmetries present in the flat space gauge theory, only four remain once
we compactify on $Q$. For large $Q$ there is then a description of the dynamics in terms of a supersymmetric
Yang-Mills theory on $M$ with minimal supersymmetry.

An important point is that the three scalars $B$ actually become the
components of a 1-form on $Q$ because of supersymmetry.
A very simple way to see this is that if we consider the fluctuations of the three components of $A$ which are
along $Q$, then these fluctuations look like scalar fields in four dimensions. But a real scalar field in four
dimensional supersymmetry must combine with either a 2-form potential or another scalar field in order to form
a representation of the supersymmetry algebra. In this case the three components of $B$ transform as 1-forms
on $Q$ and not as 2-forms on $M$ \cite{bsa1}.
Thus the components of $A$ tangential to $Q$ have bosonic superpartners
which are a 1-form $B$ on $Q$. All fields transform in the adjoint representation of the gauge group.

One can check explicitly (see appendix) that the conditions for unbroken supersymmetry imply that
\be
F(A) = B\wedge B
\ee
and
\be
D_A B = 0
\ee
ie that the field strength of the gauge field components along $Q$ are equal to the
commutator of the $B$'s and that the covariant derivative of $B$ with respect to $A$ is
zero.

These equations are the critical points of the following complex functional
\be
\omega \equiv \int_Q tr  (A+iB)\wedge d(A+iB) + {2 \over 3} (A+iB)\wedge (A+iB) \wedge (A+iB)
\ee
with respect to the complex gauge field $A+iB$. $\omega$ is the complex Chern-Simons
functional. Its critical points are complex flat connections. These are equivalent to a representation
of the fundamental group of $Q$ in the complexification of the ADE gauge group $G^{\bf C}$ modulo
conjugation.

This implies that the Chern-Simons functional $\omega$ 
is the contribution to the superpotential of the four dimensional theory
from fluctuations of $A$ and $B$ along $Q$. $\omega$ is invariant under gauge transformations
of the complexification of the gauge group as $F$-terms should be. In fact we can use the latter fact to
rederive $\omega$:

The contribution to the four dimensional potential energy coming from the fluctuations of $A$ and $B$
along $Q$ is  given by the bosonic part of the super Yang-Mills
action on $Q$. This includes the Yang-Mills action for $A$
\be
V = \int_Q tr F(A) \wedge * F(A) + \cdots
\ee

Then we can ask, what is the superpotential $W$ for which
\be
V = |{\delta W \over \delta (A+iB)}|^2
\ee

Since the real Chern-Simons functional, $\omega_{\bf R}$
for $A$ (which is $\omega$ with $B=0$ ) has the property that
\be
\delta \omega_{\bf R} \propto F(A)
\ee
complex gauge invariance of the $F$-term uniquely fixes $\omega$ to be the complex
Chern-Simons functional\ft{Strictly speaking, our derivation
of $\omega$ thus far applies only in the limit of global
supersymmetry. In subsection $3.2$ we will explicitly verify
that $\omega$ is indeed the superpotential.} 

As an aside, we note that for $G_2$-compactifications which are dual to heterotic or
Type I or other IIB orientifold compactifications on Calabi-Yau threefolds, $\omega$ is the $M$ theory `dual'
of the holomorphic Chern-Simons functional which is the superpotential in these
string compactifications.

We are going to pick our background gauge field $A_o +iB_o$ to simply be a critical point of
$\omega$ ie  a flat complex gauge field.  We are also going to assume that the moduli
space of complex flat connections is zero dimensional ie that the flat connection we choose
is isolated. This means from the low energy four dimensional point of view that all the
fluctuations of the gauge field around this background are massive. If this were not the
case then the low energy theory would contain phenomenologically undesirable massless
adjoint fields and the Kahler potential above would also require modification to include these additional
light degrees of freedom.

The Chern-Simons functional $\omega$
is a topologically invariant functional. In particular, its value
on a background flat connection is a constant
\be
\omega |_{A_0 + iB_0 } \equiv c = c_1 + i c_2
\ee

In general this constant\ft{Note that if we had chosen a more general
non-flat connection, then $\omega$ will not be constant and the formulae
below become more complicated.} is complex as we have emphasised in the above formula. In particular
the real part is only well defined modulo 1 (in appropriate units) and is essentially the more familiar
real Chern-Simons invariant. Its imaginary part however can in general be any real number.
In particular $c_2$ can -  as we will demonstrate -  be large. This fact
will be important for us in the following.

Summarising the results of this subsection: the contribution of the fields supported at the singularities
to the superpotential is given by a complex number, $c$ -  the Chern-Simons invariant of a flat complex
connection.

\bigskip
\Large
\noindent
{\bf {\sf 3. Minima of the Supergravity Lagrangian.}}
\normalsize
\bigskip

Having identified the contribution to the superpotential we can add it to the $G$-flux induced
linear superpotential and study the minima of the supergravity potential. The absolute minima
of the potential are supersymmetric vacua with a negative cosmological constant. These vacua are
characterised by solutions to the following equations:
\be
{\partial W \over \partial Z_k} + {\partial K \over \partial Z_k} W = 0
\ee
With $W$ given by\ft{We have redefined $c_1$ and $c_2$ for convenience}
\be
W = {1 \over 8\pi^2}(\Sigma_j Z_j G_j + c_1 + i c_2) 
\ee
and the first derivative of the Kahler potential by \cite{beasley}
\be
{\partial K \over \partial Z_k} = {7 \over 2}i {\int_X \phi_k \wedge *\varphi \over \int_X \varphi \wedge *\varphi}
\ee

we see that minima exist if,
\be
\Sigma t^j G_j  + c_1  = 0
\ee
and
\be
G_k = {7 \over 2} {\int_X \phi_k \wedge *\varphi \over \int_X \varphi \wedge *\varphi}(\Sigma_j s_j G_j + c_2 )
\ee

\bigskip
\large
\noindent
{\bf {\sf 3.1. Solution to Vacuum Equations.}}
\normalsize
\bigskip

The equation involving the axions implies that one linear combination of axions has a fixed vev. Note that in the quantum theory this equation gets modified by the addition
of half an integer, determined by which
``theta vacuum'' is picked.

The equations for the moduli fields $s_i$ imply that in any given solution
the 4-form field $G$ is proportional to $*\varphi$, with
a coefficient $\alpha$ which depends upon the moduli. To see this explicitly, expand $*\varphi$ in
terms of the basis of harmonic 4-forms $\rho_i$ which are dual to the $\phi_i$:
\be
*\varphi = \Sigma_j u_j \rho_j
\ee
Note that the $u_j$ depend upon the moduli fields $s_i$. The $u_j$ are homogeneous of degree ${4 \over 3}$.
Then the equations above may be rewritten as
\be
G_k = {7 \over 2}{u_k \over \Sigma_j s_j u_j}(\Sigma_i s_i G_i + c_2)
\ee
Therefore,
\be
G = \alpha *\varphi
\ee
where
\be
\alpha =  {7 \over 2\Sigma_j s_j u_j}(\Sigma_i s_i G_i + c_2) = {1 \over 2Vol(X)}(\Sigma_i s_i G_i + c_2)
\ee

Since a solution of the equations implies that $*\varphi$ is proportional to an integral cohomology class,
any solution has at most one modulus - the volume. This is because changes in the moduli which preserve
the volume of $X$ will always change the direction in $H^4 (X, {\bf R})$ in which $*\varphi$ points. We will
now describe a way to solve these equations and demonstrate that the volume modulus is also fixed.

The idea is simple and we are indebted to Dominic Joyce for proposing it. We assume that our $G_2$-manifold $X$
has a $G_2$-structure $\varphi^{\prime}$ which is proportional to the flux $G$,
\be
\alpha^{\prime} *\varphi^{\prime} =  G
\ee
Then we simply rescale the moduli of this $G_2$-metric and try to solve equation $(21)$. Note that if $X$ does
not have a $G_2$-structure for which $(23)$ is true for some fixed $G$-flux,
then we will typically be able to change the flux and adjust the moduli so that it is true. Hence we
can typically find combinations of $G_2$-manifolds and fluxes for which $(23)$ is true.

Under a rescaling of the moduli,
\ba
\varphi^{\prime} \longrightarrow \lambda^3 \varphi^{\prime}\\
*\varphi^{\prime} \longrightarrow \lambda^4 *\varphi^{\prime}\\
\alpha^{\prime} \longrightarrow \lambda^{-4} \alpha^{\prime}
\ea
Thus, denoting the moduli in the $\varphi^{\prime}$ metric as $s_j^{\prime}$, in order to solve our equations
we have that
\be
s_j = \lambda^3 {s_j}^{\prime}
\ee
and require that
\be
\alpha = \lambda^{-4} \alpha^{\prime}
\ee
This gives the following equation for the scale factor
\be
\lambda^3 \Sigma_j {s_j}^{\prime} G_j + c_2 = 2\lambda^3 Vol(X^{\prime}) \alpha^{\prime}
\ee

This has a unique solution given by
\be
\lambda^3 = {c_2 \over 2Vol(X^{\prime})\alpha^{\prime} - \Sigma_j {s_j}^{\prime} G_j}
\ee
Hence this minimum of the potential is isolated. In fact it is actually clear from the conditions for unbroken
supersymmetry $(18)$ that the volume modulus will be fixed in a given solution. This is because the equations
are not invariant under a rescaling of the moduli fields. The left hand side and the first term on the
right are invariant, but the last term involving $c_2$ is not invariant.

In summary, we have shown that if $X$ admits a $G_2$-metric
which is such that $*\varphi^{\prime}$ is proportional to $G$ then there exists an isolated minimum of the
supergravity potential with negative cosmological constant. Note that $X$ must be such that the topology of
$Q$ is rich enough to admit flat connections with non-zero $c_2$. We will discuss this issue later. For the moment
we wish to outline a dual Type IIA description of the moduli fixing mechanism
\ft{We thank E. Witten for suggesting this.} which may serve to clarify
the origin of $c_2$.

\bigskip
\large
\noindent
{\bf {\sf 3.2 Type IIA dual description.}}
\normalsize
\bigskip

In addition to being dual to compactifications of the heterotic string theory, $G_2$-compactifications of
$M$ theory can also be dual to orientifolds of Type IIA theory on a Calabi-Yau threefold $Z$.
Such string vacua can be described in the following way. Consider Type IIA on a Calabi-Yau threefold $Z$.
This theory has ${\cal N}=2$ supersymmetry. If $Z$ admits a ${\bf Z_2}$ symmetry $\sigma$ under which
the Kahler form $K$ and holomorphic 3-form $\Omega$ behave as
\ba
K \rightarrow - K\\
\Omega \rightarrow \Omega^*
\ea
then one can define an orientifold of the theory by gauging the ${\bf Z_2}$ symmetry generated by the product
of $\sigma$ and the world-sheet parity operator $P$.  Gauging this ${\bf Z_2}$ breaks the supersymmetry to
${\cal N}=1$.

The fixed points of $Z$ under $\sigma$, if non-zero, are 3-dimensional submanifolds
$Z^{\sigma}$ of $Z$ and will be smooth if $Z$ is smooth. The string theory thus has orientifold 6-planes
with topology $Z^{\sigma}{\times}M^{3,1}$, with $M$ four-dimensional spacetime. Consequently, the theory
also has a tadpole for the RR 7-form potential $C_7$, suported on the orientifold planes.
The tadpoles may be cancelled
by introducting $D$6-branes wrapping 3-dimensional submanifolds $Q_a$ of $Z$.
If $N^a$ denotes the number of branes wrapping $Q_a$, the homology classes of the $Q_a$'s must be such that
\be
\Sigma_a N^a [Q_a] = 4 [Z^{\sigma}]
\ee

This is the tadpole cancellation condition. Some examples of such vacua have been described in \cite{orient}.

If all the $Q_a$ are special Lagrangian 3-manifolds (ie supersymmetric cycles) then the configuration
of branes and orientifolds will be supersymmetric. Since such Type IIA vacua have only the metric,
dilaton and RR 1-form potential (dual to $C_7$) as the only non-zero fields, they arise in the $M$ theory
limit solely from the eleven dimensional metric. Supersymmetry then implies that the $M$ theory
limit is a compactification on a $G_2$-manifold. For instance, the $M$ theory limit of $N^a$ D6-branes
on $Q_a$ is an $A_{N_a -1}$-singularity over $Q_a$ in a $G_2$-manifold.

The description of the fields  on the D6-branes at low energies
is precisely the 7-dimensional super Yang-Mills theory we discussed in the $M$ theory limit.
The gauge field $A$ on $Q$ corresponds to the usual gauge field of the D6-brane and the
1-form $B$ on $Q$ has three components which locally represent normal fluctuations of
the brane inside the Calabi-Yau $Z$. The superpotential from the D6-brane sector is again
given by the complex Chern-Simons functional, but now we can interpret contributions to its
imaginary part as beign associated with non-zero values for the normal bundle fields $B$, since
the imaginary part vanishes if $B$ vanishes.

What kind of brane physics is associated with non-zero normal bundle fields? One natural
guess is the Myers effect \cite{myers}.
For example, a non-zero value for
\be
tr B\wedge B\wedge B
\ee
on a Dp-brane induces a coupling to the RR (p+4)-form field strength $F$ of the form
\be
trB^1 B^2 B^3 . F + permutations
\ee
where the $(p+4)$-form $F$ is contracted with the three components of $B$ to give a $(p+1)$-form which
we may integrate over the $Dp$-brane world-volume.  Since the $(p+4)$-form measures the charge
distribution of $D(p+2)$-branes in Type IIA/B theory, this coupling is associated with $Dp$-branes
which have ``expanded into'' $D(p+2)$-branes. In our context $p=6$ and the non-zero value of
$(34)$ occurs when $c_2 \neq 0$. Hence the $D6$-branes of the orientifold compactification
expand into $D8$-branes.

\bigskip
\large
\noindent
{\bf {\sf 3.2 The Potential.}}
\normalsize
\bigskip

Up to now, we have described the conditions for unbroken supersymmetry and their solutions.
Also of interest is the supergravity potential, $V$.
$V$ is determined by the standard formula
\be
V = e^K (g^{i\bar{j}} ({\partial W \over \partial Z_i} + {\partial K \over \partial Z_i} W)
({\partial \bar{W} \over \partial \bar{Z_{\bar{j}}}} + {\partial K \over \partial \bar{Z_{\bar{j}}}} \bar{W} )
-3|W|^2)
\ee
This turns out to be given by
\be
V = {\pi^2 \over 2} ( {\int_X G_X \wedge *G_X \over Vol(X)^2} +
{{(\int_X {C \over 2}}\wedge G_X + c_1 )^2 \over Vol(X)^3 } + {(c_2)^2 \over Vol(X)^3 }
+ {c_2  \int_X \Phi \wedge G_X \over Vol(X)^3 })
\ee

Note that $c_1$ and $\int_X {C \over 2}\wedge G$ appear on an equal footing in the effective
potential. We can see this explicitly from the $C$-field equations of motion in the presence of the
$ADE$-singularity supported on $Y = Q \times M^{3,1}$. In the presence of such an $ADE$ singularity
the $M$ theory Lagrangian has a contribution coupling the gauge fields on $Y$ to the $C$-field:
\be
\Delta L = \delta^4_{Y} \wedge \omega_3 (A) \wedge G
\ee
where $\omega_3$ is the Chern-Simons 3-form as a functional of the gauge field and $\delta^4_Y$
is a 4-form delta-function supported on $Y$. This means that the $C$-field equation of motion is
\be
d*G + {1 \over 2} G\wedge G + \delta^4_Y \wedge tr F\wedge F = 0
\ee
This implies that the gauge field makes a contribution to the conserved Page charge, $P$, defined by \cite{page}
\be
P = \int_X (*G + {1 \over 2} C\wedge G)/{4\pi^2}  + {c_1}
\ee
Hence,
turning on $A$ is equivalent to shifting the Page charge by the Chern-Simons invariant $c_1$.
We can also use this analysis to reconfirm our calculation of $W$ using the fact that $BPS$ domain walls in the
supergravity theory have a tension which is porportional to 
the change in $W$ as the wall is crossed.

Consider an $M2$-brane whose world-volume
is ${q}\times {\bf R}^{2,1}$ $\subset$ $Q\times {\bf R}^{3,1}$. From the effective field theory point of view this
domain wall in four dimensions can be described as an instanton on $Q\times {\bf R}$ for the gauge
field $A$. Here ${\bf R}$ is the direction transverse to the wall. The reason for this is precisely the equation of
motion $(39)$ which implies that gauge field configurations of non-zero instanton number have non-zero
membrane charge\ft{In the Type IIA description the instanton is a D2-brane inside the D6-brane.}.

To specify the instanton field requires boundary conditions
at $\pm \cal{1}$ which consist of two copies of $Q$. The requisite boundary conditions are a flat connection on
each copy of $Q$. The $M2$-brane domain wall is then an instanton which interpolates
between the two flat connections at the two ends. The tension of such a domain wall is then given by evaluating the
Yang-Mills action on $Q{\times}{\bf R}$
\be
T = \int_{Q \times {\bf R}} tr F\wedge *F = \int_{Q \times {\bf R}} trF\wedge F = c_1^+ - c_1^- \equiv \Delta c_1
\ee
ie is the difference in the Chern-Simons invariants of the two flat connections. Since the supersymmetry
algebra implies that
\be
T  = \Delta ({e^{K \over 2}}W)
\ee
we find (by comparing with the formulae for $V$) that $W$ receives a contribution which is directly
propotional to $c_1$, since $e^K \propto Vol(X)^{-3}$. Holomorphy ie the fact that the gauge field
on $Q$ naturally gets complexified, then implies the $c_2$ contribution. This completes the derivation of $W$ that we
have used throughout this paper.

An open question is
what kinds of domain walls in the theory shift the value of $c_2$ ??  The answer appears to
be domain walls which couple only to the gravitational field. This is because the $C$-field equation of
motion does not detect changes in $c_2$ and therefore the Einstein equations presumably do.

\bigskip
\large
\noindent
{\bf {\sf 3.4. The One Modulus Case.}}
\normalsize
\bigskip

We have shown above that the equations formally have an isolated minimum.
However, for consistency, we must also show that the minimum
exists in a region of field space where the supergravity approximation is valid, since this after all is where we
obtained the Kahler potential. This requires showing that the minimum exists at values of the $s_j$ for which
the size of $X$ is large compared to the Planck length.
Although this is a straightforward exercise in the general case
we will restrict our attention
so the very simple case of $G_2$-manifolds with one modulus, $s$. In this case we can be very explicit
about the dependence of the equations on $s$ and the formulae are a little less abstract.

With one modulus, homogeneity of $e^K$ fixes the Kahler potential to be
\be
K = -ln({1 \over 14\pi^2}s^7)
\ee

Then the vacuum equations assert that
\be
G = {7 \over 2s} (G s + c_2)
\ee

This has the unique solution
\be
s = -{7 \over 5}{c_2 \over G}
\ee
Note that, since $s$ is positive in the supergravity approximation, the signs of $c_2$ and $G$ must be opposite.
Since $c_2$ is fixed by the topology there is a solution for only one sign of $G$.

The supergravity approximation is valid when the volume of $X$ is large and this corresponds to $s$ being
large. Therefore as long as $c_2$ is large compared to $G$, the minimum exists in a region of field
space within the approximation. In $M$ theory $s$ has length dimension three, so a $c_2$ of order
$10^3$ or bigger and a minimal flux quantum is presumably enough to put the minimum in the regime of validity.
Before we discuss in more detail the nature of $c_2$ and the existence of $G_2$-manifolds with
non-zero $c_2$ we can compute using the supergravity potential the value of the potential in the minimum.
This is the cosmological constant of the anti-de Sitter space. This turns out to be 
\be
V_0 = \Lambda_0 = {21 \pi^2 5^5  \over 2 \times 7^5} ({G \over c_2})^5 G^2 \approx  20 {G^7 \over c_2^5}
\ee
which if $G$ is minimal is small for large $c_2$.

The potential also has another critical point at finite $s$. This critical point is a de Sitter maximum.
The cosmological constant of
de Sitter space is of the same order as it is in the minimum.
The multi modulus case also presumably has a de Sitter maximum. This follows simply from the fact that,
as we have seen, the potential has a minimum below zero at some finite value of the $s$-fields.
But for extremely large values of the fields the terms involving $c_2$ are negligible compared to the flux induced terms calculated in \cite{beasley}. These terms are purely
positive and for large $s_i$ the potential approaches zero from above. Hence, presuming the potential
does not blow up at finite distance,
it must have a maximum
above zero. In the multi-modulus case we actually expect the existence of de Sitter minima as well
although we have not been able to prove this.  This is simply because there are in general many fields.
If this is the case, the theory would have non-supersymmetric
metastable vacua with positive cosmological constant. Aspects of cosmology of these compactifications are being
investigated in \cite{ukstring}.

We now turn to the question of the existence of $G_2$-manifolds with the properties required for this
moduli fixing mechanism to be implemented.

\bigskip
\large
\noindent
{\bf {\sf 4. On the Existence of $G_2$-manifolds with $c_2 \neq 0$.}}
\normalsize
\bigskip

Recall that our $G_2$-manifold $X$ has an $ADE$-singularity supported along a 3-manifold $Q$.
$c_2$ is the imaginary part of the Chern-Simons invariant of the flat complex $ADE$-connection
on $Q$. In particular if $B$, the imaginary part of the gauge field is gauge equivalent to zero $c_2$
will vanish. For example, consider the case when the gauge group is $G^{\bf C} = SL(N, {\bf C})$.
The flat connection is then a set of matrices in $G^{\bf C}$ which generate a group which
is homomorphic to ${\cal{\pi}}_1 (Q)$.
If, by conjugation by elements in $G^{\bf C}$, one can unitarise these matrices ie conjugate
them into $G^{\bf R}$ $=SU(n)$, $B$ is gauge equivalent to zero and $c_2$ vanishes. This is certainly the
case when ${\cal{\pi}}_1 (Q)$ is finite. We should therefore consider the case of infinite fundamental
group.

As far as we are aware, the only known examples of flat connections with non-zero $c_2$ are in the
cases in which,
\be
Q = {\bf H^3}/{\Gamma}
\ee
ie when $Q$ is diffeomorphic to a compact 3-manifold which admits a hyperbolic metric\ft{In \cite{curio},
a link between $G_2$-manifolds and hyperbolic geometry
was also made.}. A hyperbolic
metric is a metric of constant negative curvature ie is locally isometric to Euclidean anti-de Sitter space.

Now, since
\be
{\bf H^3} =  SO(1,3)/SO(3) = PSL(2, {\bf C}) / PSU(2)
\ee

$\Gamma$ is therefore a subgroup of $PSL(2, {\bf C})$ and canonically defines a representation of
${\cal{\pi}}_1 (Q)$ in $PSL(2, {\bf C})$ ie a flat $PSL(2, {\bf C})$ connection on $Q$. This representation
can always be lifted to a representation of $SL(2, {\bf C})$.
This is the flat connection that we will consider using, by
regarding $SL(2, {\bf C})$ as a subgroup of $G^{\bf C}$.
The coresponding gauge field in the
standard hyperbolic metric is given by

\ba
A + iB = \left(\begin{array}{c}  f\;\;\;\; \; g \\  \;h\; -f \end{array}\right)
\ea

where $f,g,h$ are complex one forms given in terms of the dreibein $e_i$ and spin connection $\omega_{ij}$ by
\cite{calegari}
\ba
2f = ie_1 - \omega_{23}\\
2g = i(e_2 + \omega_{12}) +  e_3 + \omega_{13} \\
2h = i(e_2 - \omega_{12}) -  e_3 + \omega_{13}
\ea
Then, one can readily compute $c$ the Chern-Simons invariant in the hyperbolic metric to learn that
\be
c_2 \propto Vol({\bf H^3}/ \Gamma)
\ee

So in this case $c_2$ is the volume of
$Q$ computed in its canonically normalised hyperbolic metric. $c_2$ can in fact be  large in practise.
To see this, consider a hyperbolic 3-manifold $Q^{\prime}$ with a volume invariant $V^{\prime}$.
Then $Q^{\prime}$ admits a finite cover $Q$ whose volume invariant is
$NV^{\prime}$. $N$ can be an integer of arbitrarily high order!

The existence of such a ${\Gamma}$ implies that $\Gamma$
contains a normal subgroup of index $N$. Equivalently,
$\Gamma$ is homomorphic to a finite group of order $N$.

To show this recall that $\Gamma$ is finitely generated by
determinant one complex matrices (modulo $\pm 1$).
The entries of these matrices are complex numbers. Let ${\bf F}$ $\subset {\bf C}$
denote the sub-ring of the complex numbers generated by the entries of these matrices. ${\bf F}$
is also finitely generated. In fact $\Gamma$ is conjugate
to a representation in which the entries of the matrices
are all algebraic\ft{ie
contained in a finite algebraic extension of the
field of rational numbers.}\cite{Thurston}. 
This implies that $\Gamma$
is conjugate to $PSL(2, {\bf F})$, for some ${\bf F}$.

For instance ${\bf F}$ might be a subset of ${\bf C}$ of the form $\{ a + i b | a,b \subset {\bf Z} \}$.
We can then consider the subrings of ${\bf F}$ called ``prime ideals'' which we denote by ${\bf P}$.
For example, ${\bf P}$ might 
be $\{ cp + i dp | c,d \subset {\bf Z}\}$ with $p$ a prime number.
Then ${\bf F}/{\bf P}$ is a finite field and we have  maps
\be
\Gamma \longrightarrow PSL(2, {\bf F}) \longrightarrow PSL(2, {\bf F}/{\bf P} )
\ee

This gives a map from $\Gamma$ to a finite group
$PSL(2, {\bf F}/{\bf P})$
which by choosing the ``prime
number'' in ${\bf P}$ is of arbitrarily large order. The kernel of this map is then the fundamental group of
a hyperbolic 3-manifold whose volume is of arbitrary order.

Note that if a $G_2$-manifold has an $ADE$-singularity supported along a 3-manifold $Q$ of
hyperbolic type then the metric induced on $Q$ will typically not be the constant curvature metric.
However, a flat connection with the same Chern-Simons invariant exists on $Q$ regardless of
the metric induced upon it. Therefore $c_2$ is not the actual volume of $Q$ which in general
depends upon the $s_j$.  However, because $Q$ is a supersymmetric 3-cycle its volume
form is given by $\varphi$. Hence the volume of $Q$ is linear in the $s_j$.   Formula $(45)$ (and its generalisation to
the multi-modulus case) then
shows that the actual volume of $Q$ and its volume in the hyperbolic metric are linearly
related.

Another important point about the canonical flat connection on a hyperbolic 3-manifold is that
it is rigid ie that it admits no deformations preserving flatness. This was another property that
we were assuming in the calculation of the minimum of the effective potential. This property follows
from the uniqueness property of hyperbolic 3-manifolds called Mostow Rigidity.
The algebraic version of Mostow rigidity asserts that any two isomorphic subgroups
of $PSL(2, {\bf C})$ which occur as the fundamental
groups of finite volume hyperbolic three-manifolds
are conjugate. This implies that the flat connection is unique\ft{The proof of Mostow rigidity is somewhat outside
the scope of this paper.}.

In summary, we have seen that if the $G_2$-manifold has an $ADE$-singularity supported along
a 3-manifold $Q$ which admits a hyperbolic metric then there exists an isolated connection with non-zero
$c_2$. Moreover $c_2$ can be a large number. Do such $G_2$-manifolds exist?

An important result which emerged from Thurston's work on 3-manifolds is that ``most'' compact 3-manifolds
are of hyperbolic type. The Thurston program \cite{Thurston}
resulted in a picture of a 3-manifold in which it can be
decomposed into a collection of `prime pieces' which consist of other 3-manifolds. Each prime 3-manifold
then admits one of eight types of locally homogeneous metrics. Seven of these eight types are 3-manifolds
which are classified. For example they include finite quotients of $S^3$ and $T^3$.
The remaining geometry, namely hyperbolic 3-space, makes up all of the remaining prime 3-manifolds
and since the list of the other 7-types is so small, hyperbolic 3-manifolds make up ``most'' prime 3-manifolds.

A more concrete argument pertaining to the existence of hyperbolic submanifolds of $G_2$-manifolds is perhaps
the following. From various string duality \cite{bsa3,ew2}
and mathematical arguments \cite{dominic}, one expectation about some $G_2$-manifolds
is that they admit $K3$-fibrations over ${\bf S^3}$ (or a quotient). Imagine that the generic $K3$-fiber has not one
but several $ADE$-singularities of the same type. For example three $A_2$-singularities. Then locally, each such
singularity, as we follow it around on ${\bf S^3}$, gives a section of the fibration. However, along a codimension
two locus $L$ in ${\bf S^3}$ the three singularities of the $K3$-fibers above $L$ could meet. In this way the
$K3$-fibration would not have a section but an order 3 multi-section. This produces in the total space of $X$
a single $A_2$-singularity supported on a 3-manifold $\Sigma$. $\Sigma$
is in fact a 3-fold branched cover of ${\bf S^3}$.
An important fact about 3-manifolds is that any 3-manifold can be regarded as
a 3-fold branched cover of ${\bf S^3}$ \cite{mostenisos,birman}.
So, in this way one can imagine that $X$ is such that the branching is hyperbolic. Unfortunately, developing this
argument further requires understanding better the branching locus\ft{$X$ could have an additional singularity along $L$
at which additional light degrees of freedom might be found.} $L$.

We should also point out an analagous situation in which plenty of examples may be found. This consists
of a compact Calabi-Yau threefold $Z$ with an $ADE$-singularity localised along a hyperbolic
2-manifold ie a Riemann surface of genus $>1$.

Having given some arguments which might suggest that $G_2$-manifolds exist with non-zero $c_2$ we now
return to the physics - assuming the existence of a solution.

\bigskip
\large
\noindent
{\bf {\sf 5. Holography and AdS/CFT.}}
\normalsize
\bigskip

The alert reader may have noticed something which at first sight appears peculiar from our discussion
of the minima of the supergravity potential. This is the fact that the potential for the $s_i$ had an isolated
minimum but only one linear combination of axions had  a fixed vev in the vacuum. It thus appears that
we have an ${\cal N}$ $=1$ vacuum configuration in which $b_3(X) -1$ real scalars (the axions) are massless.
This seems peculiar at first because the usual representations of the ${\cal N}$ $=1$ supersymmetry
algebra contain two bosonic degrees  of freedom not one. As we shall explain this appears to be
related to the holographic dual description of these vacua.

The resolution of this peculiarity is that the vacuum we are talking about is a supersymmetric anti-de Sitter
vacuum. In anti-de Sitter space, in addition to supermulitplets with two bosonic degrees of freedom,
there does indeed exist a supermultiplet with one bosonic degree of freedom. This is the so called
singleton representation. Therefore, our axions must be singletons.

An important point about singleton representations is that they naturally reside on the boundary of
anti-de Sitter space\ft{See the introduction of 
\cite{singleton} for a review with detailed references
on the representation theory of the anti-de Sitter group.}. Thus, the excitations of the axion fields in our vacua represent
boundary degrees of freedom.

A very natural question to ask is what is the holographic dual description of these adS vacua?
Because of the symmetries, this dual description is almost certainly a three dimensional conformal field
theory which must - in some natural sense - reside on the boundary of anti-de Sitter space.
In the better understood adS vacua of $M$ theory, this $CFT$ is the world-volume theory of some
collection of $M$2-branes on the boundary of the spacetime.
In that context it was realised that the singleton representation in the gravity theory represented the
center of mass degree of freedom of the branes.

Since we have identified $b_3(X) -1$ axionic singleton fields in the adS vacua under discussion here,
it is natural to propose that the holographic dual description corresponds to the world-volume field
theory of $b_3(X) - 1$ collections of branes at the boundary of the spacetime (although it is not clear which kind
of branes these actually are). One could try and verify this proposal by finding the corresponding
brane solution of $M$ theory on $X$.

\bigskip
\large
\noindent
{\bf {\sf 6. Phenomenology.}}
\normalsize
\bigskip

Presuming that compact $G_2$-manifolds with non-zero $c_2$ exist, we have shown that vacua of $M$ theory
exist with no moduli. This means that such $G_2$-manifolds would provide solutions of the eleven dimensional
theory which describe four dimensional universes with no undetermined parameters except the fundamental Planck
scale.
In principle therefore all
masses and couplings are calculable by the theory in the isolated vacuum state.

The main phenomenological difficulties with the vacua we have described here are a) they are supersymmetric
and b) they have a negative cosmological constant. However, as we have seen, the moduli fixing mechanism
that we have described here can in principle be implemented in a $G_2$-manifold whose associated low
energy physics contains no light charged matter, since we did not have to consider the matter fields supported
at ``chiral singularities'' of $X$. We should therefore consider the utility of the mechanism in compactifications
which include quarks and leptons.

The basic idea that we consider is that the moduli fixing
mechanism considered here is the dominant 
contribution to the moduli potential in a more realistic
scenario. In order to investigate this possibility, it
seems necessary to assume that
in addition to the
singularities which produce a supersymmetric standard model sector, $X$ also has singularities which
produce a supersymmetry breaking sector. Once supersymmetry is broken (at a scale $m_{susy}$)
the zero point energies of
the charged fields will generate a contribution $\Lambda_1$ to the
cosmological constant of order $m_{susy}^4$ and we
will simply consider the consequences of requiring that
$\Lambda_1$ and the contribution $\Lambda_0$
of section three are of the same order. This is necessary
if one is to obtain a positive cosmological constant.
Under the {\it strong} assumption that a suitable supersymmetry breaking
mechanism exists for which this is true, there are a couple
of quite striking consequences.

First, the heirarchy problem reduces to explaining why a topological invariant $c_2$ takes
the value that it does. Second, the fundamental Planck scale is roughly 10 TeV, if we take $m_{susy}=$ 1 Tev. Thus, if
these sorts of models actually existed, they would provide
a concrete realisation of the ideas proposed in \cite{large}.

Note that the cosmological constant problem in this context then reduces to the problem of explaining why
$\Lambda_1$ cancels $\Lambda_0$ to at least $10^{56}$ orders of magnitude. 
We now turn to an explanation of these
statements.

Once supersymmetry is broken, the mass splittings between fermions and bosons will generate a contribution
to the cosmological constant of the form $m_{susy}^4$, where $m_{susy}$ is the scale of supersymmetry breaking.
If this contribution to $\Lambda$ is going to be comparable
to the cosmological constant calculated
in section three, we require that
\be
20 {G^7 \over c_2^5} m_p^4 \approx m_{susy}^4
\ee
where $m_p$ is the four dimensional Planck
mass. Thus,  the heirarchy problem essentially reduces to explaining why
\be
{c_2^5 \over G^7} \approx 20{({m_p \over m_{susy}})^4}
\ee
The minimal value for $G$ is $2\pi$, so the smallest $c_2$ can be is around $10^{13} - 10^{14}$, if we assume that
$1Tev \leq m_{susy}
\leq  10Tev$. 
So in this scenario, the
heirarchy problem reduces to explaining why a certain topological invariant is $10^{13}$ (or bigger if the
fluxes are larger).

On the other hand, $m_p$ is determined in terms of the eleven dimensional Planck mass, $M_p$ through
the formula
\be
m_p^2 \approx ({c_2 \over G})^{7/3} M_p^2
\ee

Hence, assuming that $m_{susy}$ is around the $TeV$ scale and we know $m_p$, we can determine the fundamental
Planck scale. For the minimal value of $c_2$, we find that the fundamental scale is one  order larger than
$m_{susy}$
\be
M_p \approx (10 - 100) TeV \approx 10  m_{susy}
\ee

For a much larger $G$-flux, say of order $10^{10}$,
\be
c_2 \approx 10^{27}
\ee
and
\be
M_p \approx 1 GeV
\ee

so vacua in which $c_2$ and $G$ are vastly different, have fundamental scales which differ by a much
milder amount. However, a fundamental scale which is lower
than the scales already probed by particle physics experiments
is surely problematic and hence models with flux-quanta
which are too large are ruled out.

We should also compute the masses of Kaluza-Klein particles. These are given by
\be
m_{kk} \approx ({G \over c_2})^{1/3}M_p
\ee
For minimal $G$-flux, this is roughly 10 GeV. For larger fluxes these masses are smaller.

An important comment is in order. If $X$ is isotropic in the minimum all its moduli are of the same order
of magnitude in the minimum of the potential. 
Naively this would imply that these vacua are ruled out
by accelerator experiments, because Kaluza-Klein excitations
of the gauge fields along the $Q_a$ which supports the
standard model gauge group would produce charged
particles of mass 10 GeV or less. Renormalisable
couplings between these particles and those of the standard
model would then imply that such isotropic vacua are inconsistent else the Kaluza-Klein modes would have been
observed in particle physics experiments above 10 GeV.

However, recall that the standard model particles are
supported at singular points on $X$ and therefore transform
only under four-dimensional gauge transformations, whereas
the gauge fields transform under {\it seven} dimensional
gauge transformations. The charged Kaluza-Klein modes 
therefore
transform under seven dimensional gauge transformations.
Because of this, {\it renormalisable}
interactions between standard model particles
and the Kaluza-Klein modes are difficult to imagine and hence
the interactions of these modes will presumably be much harder to detect.

Even if the charged Kaluza-Klein modes do have conventional
couplings to quarks and leptons, many models could
exist in which their masses are 100 GeV or much larger.
To see this, we consider the generalisation of $(45)$ to
the general case,

\be
\Sigma_i s^i G_i = -{7 c_2 \over 5}
\ee

The $Q$'s which support the gauge fields have volumes which are linear in the $s^i$ because they are
supersymmetric cycles. Hence if $s^1$ is the volume of one of the $Q$'s the above formula asserts that
as $G_1$ is increased, $s^1$ decreases. So a larger flux in one particular direction in the cohomology of $X$
corresponds to shrinking a 3-manifold in $X$. Therefore, the $Q$ which supports the standard model gauge
group can easily be much smaller than the largest scale of $X$ and if the mass scale of $Q$ is 100 GeV or
bigger the model is consistent with accelerator experiments. These smaller 3-cycles make a negligible
contribution to the volume of $X$. This implies that the cosmological constant of section three
is still given by  $(46)$ , but with the $G$ which appears there the smallest component of $G$-flux
in $X$.

It would certainly be of interest to develop some of the ideas
presented here further. In particular, a construction of
$G_2$-manifolds with some of the properties described here
would be an important step.

\bigskip
\large
\noindent
{\bf {\sf Appendix: Supersymmetry.}}
\normalsize
\bigskip

A relatively efficient method to derive the equations for unbroken supersymmetry on the fields
supported at the $ADE$-singularity along $Q$ is actually in a dual context. Namely when
$Q$ is a supersymmetric 3-cycle in a Calabi-Yau threefold $Z$ in Type IIA and some
number of $D$6-branes are wrapping $Q$. Then, as shown in \cite{bsa1}, the supersymmetry
conditions for the world-volume fields are identical to those in $M$ theory. This is because the
strong coupling limit of this Type IIA configuration is precisely an $ADE$-singularity along $Q$.

In flat space the world-volume theory of $N$ $D$6-branes is $U(N)$ super Yang-Mills in seven dimensions.
We can derive the conditions for unbroken supersymmetry by dimensional reduction of the corresponding
conditions in ten dimensional super Yang-Mills theory on a $D$9-brane.
In curved space, the analagous procedure is to dimensionally reduce the conditions for unbroken
supersymmetry for gauge fields along $Z$ to $Q$. Namely, we should consider the dimensional
reduction of the Hermitian Yang-Mills equations on $Z$ onto $Q$.

In components in a locally flat frame, the Hermitian Yang-Mills equations are linear equations on
the field strength $F_{AB}$ of a gauge field $a_B$ on $Z$:
\ba
F_{12} + F_{34} + F_{56} = 0\\
F_{13} - F_{24} = 0\\
F_{14} + F_{23} = 0\\
F_{15} - F_{26} = 0\\
F_{16} + F_{25} = 0\\
F_{35} - F_{46} = 0\\
F_{36} + F_{45} = 0\\
\ea

The first equation depends upon the metric on $Z$ via the Kahler form which we have chosen to be
\be
\omega =  dx_1 \wedge dx_2 + dx_3 \wedge dx_4 + dx_5 \wedge dx_6
\ee

The remaining six equations are the conditions that the gauge field is holomorphic. These do not depend
upon the Kahler metric. These equations correspond to $F$-terms in four dimensions and are the critical points of
the holomorphic Chern-Simons functional.

$Q$ is  a supersymmetric 3-cycle in $Z$ and as such
\be
\omega|_Q = 0
\ee

So in the local frame $e_i$ we will choose the frames along $Q$ to be spanned by $(e_1, e_3, e_5 )$.
The ${\bf R^3}$-subspace of ${\bf C^3}$ thus defined is indeed special Lagrangian with the
complex structure defined by $\omega$.  Correspondingly we will
define the gauge field components tangential and normal to $Q$ by
\ba
(a_1 ,  a_3 , a_5) \equiv (A_1 , A_2 , A_3)\\
(a_2 , a_4 , a_6) \equiv (B_1 , B_2 , B_3)
\ea

Dimensionally reducing the Hermitian Yang-Mills equations onto $Q$ then corresponds to
setting to zero the normal derivatives:
\be
\partial_2 = \partial_4 = \partial_6 = 0
\ee

If we regard both $A_i$ and $B_i$ as components of a 1-form on $Q$ (with values in the Lie algebra of the
gauge group) then the above seven equations become:
\ba
d*B + A\wedge *B + *B\wedge A = 0\\
dA + A\wedge A = B\wedge B\\
dB + A\wedge B - B\wedge A = 0
\ea

The first of these depends upon the metric on $Q$ and corresponds in four dimensions to a $D$-term.
The remaining six equations do not depend upon the metric and are $F$-terms. They are the critical
points of a complex Chern-Simons action, as described in the text.

\bigskip
\large
\noindent
{\bf {\sf Acknowledgements.}}
\normalsize

We would like to thank M. Atiyah, T. Banks, M. Douglas, N. Dunfield, 
F. Denef, D. Friedan, D. Joyce, 
S. Kachru, N. Lambert, D. Long, J. Maldacena, G. Moore,
E. Silverstein and E. Witten for very useful discussions.

\end{document}